\documentclass[12pt,superscriptaddress,preprintnumbers,showpacs,citesort]{revtex4}
\usepackage{amssymb,amsmath}
\begin{document}

\title {Local unitary equivalent consistence for $n$-party states and their
$n-1$-party reduced density matrices}

\author{WANG Zhen }
\affiliation {Department of Mathematics, Jining University,
Qufu 273155}
\author{WANG He-Ping}
\affiliation {School of Mathematical Sciences, Capital Normal
University, Beijing 100048}
\author{WANG Zhi-Xi}
\affiliation {School of Mathematical Sciences, Capital Normal
University, Beijing 100048}
\author{FEI Shao-Ming}
\affiliation {School of Mathematical Sciences, Capital Normal
University, Beijing 100048}

\begin{abstract}
We present that the
local unitary equivalence of $n$-party pure states is consistent with the one
of their ($n-1$)-party reduced density matrices. As an application,
we obtain the local invariants for a class of tripartite pure qudits.
\end{abstract}

\maketitle

As the properties of entanglement for $n$-party quantum states remain invariant under local
unitary transformations on subsystems, the entanglement can be characterized in principle
by all invariants under local transformations \cite{rains}. The polynomial invariants of local unitary
transformations have been discussed \cite{werner,sudbery,linden}. For two bipartite pure states
the set of Schmidt coefficients forms a complete set of invariants under local unitary transformations
\cite{schmidt}. For multipartite pure and mixed states, complete sets of invariants have not
been found except some particular cases, e.g. two-qubit mixed states \cite{makhlin}, three-qubit pure states
\cite{linden3,sun1} and generic mixed states \cite{albererio2,albererio3,sun}.
S. Albeverio {\it et.al.} \cite{albererio1} introduced the notion of generic states and obtained criteria for local equivalence of such states. In  \cite{albererio4} they further introduce the concept of CHG (commuting  high generic) states maintaining the criteria of local equivalence and characterize the equivalence classes under local unitary transformations for the set of tripartite states whose partial trace with respect to one of the subsystems belongs to the class of CHG mixed states.
In this letter we show that the equivalent problem
under local unitary transformations for any $n$-party
pure state can be reduced to the problem for its $(n-1)$-party reduced states.
According to the result, one can easily obtain the set of invariants for $n$-party pure states with
the known invariants for $(n-1)$-party mixed states.

Let $W_j$ be a $d_j\times d_j$ unitary matrix, and the subsystem ${\cal H}_{A_j}$ be
a Hilbert space of $d_j$ dimension.
We first give a lemma which can be proved as in \cite{nilsen} (p.111, Exercise 2.80).

{\bf Lemma} \quad If two $n$-party pure states $|\psi\rangle$ and
$|\psi^\prime\rangle$ acting on the system ${\cal H}_{A_1}\otimes {\cal
H}_{A_2}\otimes\cdots\otimes{\cal H}_{A_n}$ have same ($n-1$)-party
reduced density matrices, then for each party $j\in\{1,\ 2,\ \cdots,\ n\}$,
there is a unitary transformation $W_j$ such that
$|\psi^\prime\rangle=(I\otimes\cdots\otimes I\otimes W_j\otimes
I\otimes\cdots\otimes I)|\psi\rangle$, where $W_j$ acts on the subsystem
of party $j$.

{\bf Remark 1}\quad The lemma shows that two pure states with the same reduced matrices are equivalent under
local unitary transformations. However, it is not true for mixed states. For example, let
\begin{eqnarray}
\rho_1&=&\frac{1}{3}|\psi^+\rangle\langle\psi^+|+\frac{2}{3}|\psi^-\rangle\langle\psi^-|, \label{ghz1}\\
\rho_2&=&\frac{1}{2}|\psi^+\rangle\langle\psi^+|+\frac{1}{2}|\psi^-\rangle\langle\psi^-|,
\label{ghz2}
\end{eqnarray}
where
\begin{displaymath}
|\psi^\pm\rangle=\frac{1}{\sqrt{2}}(|000\rangle\pm|111\rangle).
\end{displaymath}
It is obvious that
${\rm Tr}_{A_1}(\rho_1)={\rm
Tr}_{A_1}(\rho_2),\ {\rm Tr}_{A_2}(\rho_1)={\rm
Tr}_{A_2}(\rho_2)$ and ${\rm Tr}_{A_3}(\rho_1)={\rm
Tr}_{A_3}(\rho_2)$.
However, there is no local unitary
transformation $W_j$ such that $\rho_1=(I\otimes\cdots\otimes
I\otimes W_j\otimes I\otimes\cdots\otimes
I)\rho_2(I\otimes\cdots\otimes I\otimes W_j\otimes
I\otimes\cdots\otimes I)^\dagger$ for any party $j$ as the
ranks of $\rho_1$ and $\rho_2$ are different.

One may ask that if the reduced matrices of two pure
multipartite states are equivalent under local unitary
transformations, whether the two pure multipartite states are
equivalent under local unitary transformations? The answer is affirmative:

{\bf Proposition}\quad If one of $(n-1)$-party reduced density
matrices for $n$-party pure states $|\psi\rangle$ is local unitary
equivalent to the corresponding $(n-1)$-party reduced density matrix
for $n$-party pure states $|\psi^\prime\rangle$ in the system ${\cal
H}_{A_1}\otimes {\cal H}_{A_2}\otimes\cdots\otimes{\cal H}_{A_n}$,
the two $n$-party pure states $|\psi\rangle$ and $|\psi^\prime\rangle$ are
also local unitary equivalent.

{\bf Proof.}\quad Let $j\in\{1,\ 2,\ \cdots,\ n\}$ be a party label.
Set $\rho_{(j)}$ and $\rho_{(j)}^\prime$ be the $(n-1)$-party
reduced density matrix obtained
by taking the partial trace of $|\psi\rangle$ and $|\psi^\prime\rangle$ over party $j$,
respectively. In addition,
$\rho_{(j)}$ and $\rho_{(j)}^\prime$ are equivalent under local unitary transformations.
Hence there exist unitary operators $U_1,\ \cdots, U_{j-1},\ U_{j+1},\ \cdots,\ \ U_n$ such
that $\rho_{(j)}^\prime=(U_1\otimes\cdots\otimes U_{j-1}\otimes
U_{j+1}\otimes\cdots\otimes U_n)\rho_{(j)} (U_1\otimes\cdots\otimes
U_{j-1}\otimes U_{j+1}\otimes\cdots\otimes U_n)^\dagger$.

Suppose $U_j$ be a unitary operator in the subsystem ${\cal H}_{A_j}$.
Let $|\phi\rangle=(U_1\otimes U_2\otimes\cdots\otimes U_n)|\psi\rangle$,
then
\begin{eqnarray}
\sigma_{(j)}&=&{\rm Tr}_j(|\phi\rangle\langle\phi|)\nonumber\\
&=&(U_1\otimes\cdots\otimes U_{j-1}\otimes
U_{j+1}\otimes\cdots\otimes U_n)\rho_{(j)}\nonumber \\
&&(U_1\otimes\cdots\otimes U_{j-1}\otimes
U_{j+1}\otimes\cdots\otimes U_n)^\dagger \nonumber\\
&=&\rho_{(j)}^\prime,\nonumber
\end{eqnarray}
where $\sigma_{(j)}$ is the $(n-1)$-party reduced density matrix of
$n$-party pure states $|\phi\rangle$ for $n-1$ subsystems except ${\cal
H}_{A_j}$. As the pure state $|\phi\rangle$ have same reduced
density matrices as $|\psi^\prime\rangle$, according to the lemma, we
obtain that there exists a unitary transformation $V_j$ such
that $|\phi\rangle=(I\otimes\cdots\otimes I\otimes V_j\otimes
I\otimes\cdots\otimes I)|\psi^\prime\rangle$, where the unitary operator
$V_j$ acts on the subsystem ${\cal H}_{A_j}$, i.e., $(U_1\otimes
U_2\otimes\cdots\otimes U_n)|\psi\rangle=(I\otimes\cdots\otimes I\otimes
V_j\otimes I\otimes\cdots\otimes I)|\psi^\prime\rangle$. Thus
$|\psi^\prime\rangle=(U_1\otimes U_2\otimes\cdots\otimes V_j^\dagger
U_j\otimes\cdots\otimes U_n)|\psi\rangle$, which proves the result.
$\hfill\Box$

{\bf Remark 2}\quad If $|\psi\rangle$ and $|\psi^\prime\rangle$ only have one same
$(n-1)$-party reduced density matrix, e.g., $\rho_{(1)}=\rho_{(1)}^\prime$,
and $\rho_{(i)}\neq\rho_{(i)}^\prime$ for any $i=2,\ 3,\ \cdots,\ n$,
then $n$-partite pure states $|\psi\rangle$ and $|\psi^\prime\rangle$ are still
equivalent under unitary transformations.

If two $n$-party pure states $|\psi\rangle$ and
$|\psi^\prime\rangle$ in the systems ${\cal H}_{A_1}\otimes {\cal
H}_{A_2}\otimes\cdots\otimes{\cal H}_{A_n}$ are equivalent under
local unitary transformations, then their reduced density matrices must
be equivalent under local unitary transformations. Thus the local
unitary equivalence of multipartite pure states is consistent with
the local unitary equivalence of their reduced density matrices.
This is, however, not the case for multipartite mixed states,
e.g. the states in (\ref{ghz1}) and (\ref{ghz2}).

The proposition yields the fact that the problem of invariants for
$n$-party pure states can be reduced to the one for ($n-1$)-party mixed states.
For example, two pure states are equivalent under local unitary transformations
if and only if they have the same values of the invariants $I_\alpha,\ \alpha=1,\ \cdots,\ n$
\cite{popescu}.

As an application, we discuss
the invariants for tripartite pure qudits in the following. Let ${\cal H}_i
(i=1,\ 2,\ 3)$ be  complex Hilbert spaces of dimension $d$. A mixed state $\rho$ in ${\cal H}_1\otimes
{\cal H}_2$ with ${\rm rank}(\rho)=n\leq d^2$ can be decomposed according to its eigenvalues $\lambda_i$
and eigenvectors $|v_i\rangle,\ i=1,\ 2,\ \cdots,\ n$:
\begin{displaymath}
\rho=\sum_{i=1}^n\lambda_i|v_i\rangle\langle v_i|,
\end{displaymath}
where $|v_i\rangle$ has the form
\begin{displaymath}
|v_i\rangle=\sum_{k,l=1}^na_{kl}^i|kl\rangle,\ a_{kl}^i\in{\cal C},\
\sum_{k,l=1}^na_{kl}^ia_{kl}^{i*}=1,\ i=1,\ 2,\ \cdots,\ n.
\end{displaymath}

Let $A_i$ denote the matrix given by $(A_i)_{kl}=a_{kl}^i$,
$\rho_i={\rm Tr}_2|v_i\rangle\langle v_i|=A_iA_i^\dagger$,
$\theta_i=({\rm Tr}_1|v_i\rangle\langle v_i|)^*
=A_i^T A_i^*,\ i=1,\ 2,\ \cdots,\ n$, where ${\rm Tr}_1$ and ${\rm Tr}_2$ stand for
the traces over the first and second Hilbert spaces, respectively. Two ``metric tensor"
matrices $\Omega(\rho)$ and $\Theta(\rho)$ is given with entries
\begin{displaymath}
\Omega(\rho)_{ij}={\rm Tr}(\rho_i\rho_j),\
\Theta(\rho)_{ij}={\rm Tr}(\theta_i\theta_j),
\end{displaymath}
for $i,\ j=1,\ 2,\ \cdots,\ n$, and
\begin{displaymath}
\Omega(\rho)_{ij}=\Theta(\rho)_{ij}=0,
\end{displaymath}
for $n<i,\ j\leq d^2$.

Let $|\psi\rangle$ be a tripartite pure qudit in ${\cal H}_1\otimes{\cal H}_2\otimes{\cal H}_3$.
It can be regarded as a bipartite state by taking ${\cal H}_1\otimes{\cal H}_2$ and ${\cal H}_3$
as the two subsystems. Then denote the bipartite decomposition of $|\psi\rangle$ as $12-3$.
Set $a_{ijk}$ be the coefficients of $|\psi\rangle$ in orthonormal bases
$|e_i\rangle\otimes|f_j\rangle\otimes|h_k\rangle$, here $|e_i\rangle,\ |f_i\rangle$ and $|h_i\rangle$ are orthonormal bases
in Hilbert spaces ${\cal H}_1,\ {\cal H}_2$ and ${\cal H}_3$, respectively. Let ${\cal A}$ denote the
matrix with respect to the bipartite state in $12-3$ decomposition, i.e. taking the subindices $ij$ and
$k$ of $a_{ijk}$ as the row and column indices of ${\cal A}$. Taking partial trace of $|\psi\rangle\langle\psi|$ over
the third subsystem, we have $\sigma={\rm Tr}_3|\psi\rangle\langle\psi|={\cal A}{\cal A}^\dagger$. The reduced density matrix
$\sigma$ can be decomposed according to their eigenvalues $\mu_i$ and eigenvectors $|\xi_i\rangle$,
\begin{equation}
\sigma=\sum_{i=1}^{n}\mu_i|\xi_i\rangle\langle\xi_i|,
\end{equation}
where $i=1,\ 2,\ \cdots,\ n$.
Let ${\cal A}_i$ denote the matrix with entries given by the coefficients of $|\xi_i\rangle$ in the bases
$|e_k\rangle\otimes|f_l\rangle$. We have
$\rho_i^\prime={\rm Tr}_2|\xi_i\rangle\langle\xi_i|={\cal A}_i{\cal A}_i^\dagger$,
$\theta_i^\prime=({\rm Tr}_1|\xi_i\rangle\langle\xi_i|)^*={\cal A}_i^T{\cal A}_i^*$, $i=1,\ 2,\ \cdots,\ n$.
Denote $\Omega^\prime(|\psi\rangle)$ and $\Theta^\prime(|\psi\rangle)$ be two matrices with entries given by
\begin{equation}\label{tomega1}
\Omega^\prime(|\psi\rangle)_{ij}={\rm Tr}(\rho^\prime_i\rho^\prime_j),\
\Theta^\prime(|\psi\rangle)_{ij}={\rm Tr}(\theta^\prime_i\theta^\prime_j),
\end{equation}
for $i,\ j=1,\ 2,\ \cdots,\ n$, and
\begin{equation}\label{tomega1}
\Omega^\prime(|\psi\rangle)_{ij}=\Theta^\prime(|\psi\rangle)_{ij}=0,
\end{equation}
for $n<i,\ j\leq d^2$.
Let ${\cal G}^\prime$ be a class of tripartite pure states $|\psi\rangle$ satisfying
\begin{equation}\label{tgeneric}
{\rm Det}(\Omega^\prime(|\psi\rangle))\neq0,\ {\rm Det}(\Theta^\prime(|\psi\rangle))\neq0.
\end{equation}

Using the proposition, we can obtain a set of invariants for
tripartite pure qudits in  ${\cal G}^\prime$:
\begin{eqnarray}\label{invariant}
&&{J^\prime}^s(|\psi\rangle)={\rm Tr}_2({\rm Tr}_1{\rho^\prime}^s),\ \ s=1,\ 2,\ \cdots,\ d^2;\nonumber \\
&&\Omega^\prime(|\psi\rangle)_{ij}={\rm Tr}(\rho^\prime_i\rho^\prime_j),\
\Theta^\prime(|\psi\rangle)_{ij}={\rm Tr}(\theta^\prime_i\theta^\prime_j),\ i,\ j=1,\ 2,\ \cdots,\ n; \nonumber \\
&&X^\prime(|\psi\rangle)_{ijk}={\rm Tr}(\rho^\prime_i\rho^\prime_j\rho^\prime_k),\ Y^\prime(|\psi\rangle)_{ijk}={\rm Tr}
(\theta^\prime_i\theta^\prime_j\theta^\prime_k),\ i,\ j,\ k=1,\ 2,\ \cdots,\ n.
\end{eqnarray}
Similarly, we can define the set of states ${\cal G}^{\prime\prime}$ (${\cal G}^{\prime\prime\prime}$) via regarding $|\psi\rangle$
as a bipartite state in ${\cal H}_1$ (${\cal H}_2$) and ${\cal H}_2\otimes{\cal H}_3$ (${\cal H}_1\otimes{\cal H}_3$).
The corresponding set of invariants for tripartite pure qudits in ${\cal G}^{\prime\prime}$ and
${\cal G}^{\prime\prime\prime}$ can be obtained in an analogous way.

It is well known that the invariants for two-qubit mixed states are studied and a complete set of
18 polynomial invariants is presented \cite{makhlin}. One can get the polynomial invariants of every three-qubit
pure state $|\psi\rangle$ by using our proposition. Unlike the invariants obtained by
Y. Makhlin in \cite{makhlin},
the polynomial invariants can be represented by the coefficients of $|\psi\rangle$ under any bases
and is convenient to calculate the invariants.

\bigskip
\noindent{\bf Acknowledgments}\, This work was supported by the NNSF of China (Grant Nos 10871227, 10875081),
the NSF of Beijing (Grant No 1092008), PHR(IHLB)201007107 and the Shandong Province Doctoral Fund(Grant No BS2010DX004).


\smallskip
\begin{thebibliography}{99}
\bibitem{rains} Rains E M 2000 IEEE Trans. Inf. Theory 46  54

\bibitem{werner} Vollbrecht K G H and Werner R F 2001 Phys. Rev. A 64  062307

\bibitem{sudbery} Sudbery A 2001  J. Phys. A: Math. Gen. 34   643 \

\bibitem{linden} Barnum  H and Linden N 2001  J. Phys. A: Math. Gen. 34   6787\


\bibitem{schmidt} Schmidt E 1906 Math. Annalen. 63  433 \

\bibitem{makhlin} Makhlin Yu 2002 Quantum Inf. Proc.  1  243 \

\bibitem{linden3} Linden N, Popescu S and Sudbery A 1999  Phys. Rev. Lett. 83 243

\bibitem{sun1} Sun B Z  and Fei S M 2006  Commun. Theor. Phys. 45  1007 \

\bibitem{albererio2} Albeverio S,  Fei S M, Parashar P and Yang W L 2003  Phys. Rev. A 68, 010303(R)

\bibitem{albererio3} Albererio S, Fei S M and  Goswami D  2005 Phys. Lett. A  340 37

\bibitem{sun} Sun B Z,  Fei S M, Jost-Li X Q and Wang Z X 2006  J.Phys.A: Math.Gen. 39 L43 \

\bibitem{albererio1} Albeverio S, Cattaneoa L, Fei S M and  Wang X H 2005  Rep. Math. Phys. 56
341\

\bibitem{albererio4} Albeverio S, Cattaneoa L, Fei S M and  Wang X H 2006  Rep. Math. Phys. 58
223\

\bibitem{nilsen} Nilsen M and Chuang I L 2000  Quantum Computing and Quantum Information
(Cambridge: Cambridge University Press)

\bibitem{popescu}  Linden N and Popescu S  1998  Fortsch. Phys.  46
567\

\end{thebibliography}
\end{document}